\documentclass[graybox]{svmult}

\usepackage{mathptmx}       
\usepackage{helvet}         
\usepackage{courier}        
\usepackage{type1cm}        
\usepackage{makeidx}         
\usepackage{graphicx}        
\usepackage{multicol}        
\usepackage[bottom]{footmisc}
\usepackage{amssymb, amsmath, amsfonts}

\DeclareMathOperator{\ocap}{\displaystyle{\small{\textcircled{{\scriptsize $\cap$}}}}}

\makeindex             

\begin{document}

\title*{Measuring the expertise of workers for crowdsourcing applications}
\author{Jean-Christophe Dubois, Laetitia Gros, Mouloud Kharoune, Yolande Le Gall, Arnaud Martin, Zoltan Miklos and Hosna Ouni}
\authorrunning{Jean-Christophe Dubois et al.}
\institute{Jean-Christophe Dubois, Mouloud Kharoune, Yolande Le Gall, Arnaud Martin, Zoltan Miklos and Hosna Ouni \at Univ Rennes, CNRS, IRISA IRISA, DRUID team, Lannion, France, \email{hosnaouni@gmail.com, (jean-christophe.dubois, mouloud.kharoune, yolande.le-gall, arnaud.martin, zoltan.miklos)@univ-rennes1.fr}
\and Laetitia Gros \at Orange Labs, Lannion, France, \email{laetitia.gros@orange.com}}
%
%
\maketitle

\abstract*{Crowdsourcing platforms enable companies to propose tasks to a large crowd of users. The workers receive a compensation for their work according to the serious of the tasks they managed to accomplish. The evaluation of the quality of responses obtained from the crowd remains one of the most important problems in this context. Several methods have been proposed to estimate the expertise level of crowd workers. We propose an innovative measure of expertise assuming that we possess a dataset with an objective comparison of the items concerned. Our method is based on the definition of four factors with the theory of belief functions. We compare our method to the Fagin distance on a dataset from a real experiment, where users have to assess the quality of some audio recordings. Then, we propose to fuse both the Fagin distance and our expertise measure.}

\abstract{Crowdsourcing platforms enable companies to propose tasks to a large crowd of users. The workers receive a compensation for their work according to the serious of the tasks they managed to accomplish. The evaluation of the quality of responses obtained from the crowd remains one of the most important problems in this context. Several methods have been proposed to estimate the expertise level of crowd workers. We propose an innovative measure of expertise assuming that we possess a dataset with an objective comparison of the items concerned. Our method is based on the definition of four factors with the theory of belief functions. We compare our method to the Fagin distance on a dataset from a real experiment, where users have to assess the quality of some audio recordings. Then, we propose to fuse both the Fagin distance and our expertise measure.}

\section{Introduction}
\label{sec:1}
Crowdsourcing was introduced by~\cite{howe2006rise}. It consists in using a collective participation to perform specific complex or time-consuming tasks that companies do not wish to carry out  internally because of a lack of resource or time. Based on sharing and collaboration, crowdsourcing belongs to the web 2.0 work framework, which enables websites users to share ideas and knowledge through dedicated platforms and websites.

Platforms like Amazon Mechanical Turk (AMT), Microworker and Foule Factory are designed to perform short tasks that computers would be unable to complete in a quick and reliable way. These tasks, such as emotion analysis, and product categorization or design comparison, are usually simple and short. 

Nevertheless, crowdsourcing platforms lead to some uncertainty, due to an uncontrolled user environment. As a result, quality assessment and reliability of contributions and workers is essential to guarantee a trouble-free process. Thus, several studies have been suggested in order to identify the experts and the most serious workers of the platform.

This paper introduces a new expertise measure using a graph distance based on the theory of belief functions. This measure enables to assess the accuracy of the workers' answers in the context of a campaign designed by Orange Labs. During this campaign, people on a crowdsourcing platform are asked to evaluate the quality of musical sequences processed by different audio encoders. 

During this 2-step study, workers have to listen to 4 HITs (\textit{Human Intelligence Task}) composed of 12 musical sequences of different qualities and evaluate their sound quality on a 5-category scale (Bad, Poor, Fair, Good, Excellent). Each category is assigned a rating from 1 (Bad) to 5 (Excellent)~\cite{ITU96}. Workers listen to the musical sequences in a random order. Only 5 of these sequences have a known quality thanks to the addition of a signal-modulated noise, with different signal-to-noise ratios (MNRUs: \textit{Modulated Noise Reference Unit}~\cite{ITU96}). These five signals, called $R_{i}$, $i=1, \ldots 5$, in the following, are considered as main references to compare the sequences inside the quality. They are expected to give quality scores from 1 (the worst) to 5 (the best). In this study, the $R_{i}$ reference signals allow to define expertise degrees. The purpose of the method is to structure the workers' answers using oriented graphs based on the 1 to 5 ratings which represent the preferences between the $R_{i}$ signals. Those graphs are then compared to the reference graph which is generated on the base of expected theoretical ratings. The results will then be used to select the experts and to focus on their performance in the 2nd step of the study. It consists in classifying the remaining 7 musical sequences in order to achieve the quality classification of these signals.

Comparing a set of graphs requires to be able to quantify the similarity between two graphs. It is a very common problem when working on social networks. Indeed, we raise this issue to understand and compare the topological properties of those graphs. Unfortunately, there is no metric or universal method for graph similarity assessment and the comparison of their geometric characteristics is an extremely complex problem. Moreover, from the algorithmic point of view, the classical methods to approach this kind of problems are complex. 

Usually, conventional listening tests are performed in laboratories. In this study concerning crowdsourcing, the answers are provided by humans in an uncontrolled environment, and it is therefore necessary to model the imperfections and undesired effects. The theory of belief functions provides a way to tackle the issue and to bring a theoretical frame to combine the pieces of information collected from different sources (workers).

In this paper, we propose an innovative approach which allows to estimate a measure of expertise using a comparison of graphs. The following section introduces the reader to the basic concepts of the theory of belief functions and then we will briefly review the existing approaches in section~\ref{sec:3}. The section~\ref{sec:4} will be structured in two parts: we will firstly present our original approach to measure an expertise degree by using a representation of answers based on graphs. Then, we will present the Fagin distance as a reference and comparison in this ranking issue. Finally, the evaluation of the method on actual data will be discussed in section~\ref{sec:5}.

\section{The theory of belief functions}
\label{sec:2}
The theory of belief functions has been introduced by~\cite{dempster1967upper} and \linebreak \cite{shafer1976mathematical}. It provides a way to represent both uncertainty and imprecision, and also to allow the ignorance of a source ({\em i.e.} a worker who can't give a response in our case). Considering a set $\Omega=\{\omega_1,\omega_2,\ldots ,\omega_n\}$ which represents the universe of possible answers to a question, a mass function is defined on $2^\Omega$ (set of all disjunctions of $\Omega$) in the $[0, 1]$ interval with the constraints: 

\begin{equation}
	\label{bba}
	\left\{
	\begin{array}{l}
	\displaystyle \sum_{A\subseteq\Omega} m(A)=1\\
	m(\emptyset)=0
	\end{array}
	\right.
\end{equation}

The mass value $m(A)$ represents the part of the belief allocated to the $A$ proposition and that can not be assigned to a strict subset of $A$. It might be seen as a family of weighted sets or as a generalized probability distribution. A set $A$ is a focal element if $m(A)\neq 0$. For example, if we consider the mass function $m(\{\omega_1, \omega_2\})=0.8$, $m(\Omega)=0.2$, this quantity represents an imprecision on $\omega_1$ or $\omega_2$ and an uncertainty because the value assigned to this proposition is 0.8.

Dealing with imperfect data from different sources requires to merge information. We therefore need to combine the mass functions in order to obtain a generic and relevant knowledge state. The conjunctive combination operator proposed by \linebreak \cite{Smets} can be computed from two mass functions from two sources through the following:

\begin{equation}
	\label{comb_smets}
	(m_{1} \ocap m_{2})(A)= \sum_{B_{1} \cap B_{2}=A}m_{1}(B_{1})m_{2}(B_{2})
\end{equation}

At the end of this combination, the mass assigned to the empty set might be interpreted as the inconsistency coming from the fusion. In order to make a decision or to define a measure, we need to evaluate the deviation from an expected mass function. Several distances have been proposed for this task. The most commonly used distance is the one from~\cite{Jousselme}. It has been adopted thanks to its properties of weight distributions as a function of the inaccuracy of the focal elements. 
It is given by:

\begin{equation}
\label{Jouss}
d_J (m_1,m_2 )=\frac{1}{2} (m_1-m_2 )^T \underline{\underline{D}} (m_1-m_2 )
\end{equation}
with: 
\begin{eqnarray}
\underline{\underline{D}}(X,Y)= \left\{
	\begin{array}{l}
1 \mbox{ if } X=Y=\emptyset\\
 \displaystyle \frac{\arrowvert X \cap Y\arrowvert}{\arrowvert X \cup Y\arrowvert} \,\, \forall X,Y \in 2^\Omega
												
	\end{array}
	\right.
\end{eqnarray}

\section{Related works for expert characterization in crowdsourcing}
\label{sec:3}

The identification of experts on crowdsourcing platforms has been the subject of several recent studies. Two different types of approach have been used: the ones where no prior knowledge is available and the ones using questions whose correct answers are known in advance. These questions with their known values are called ``golden data''\footnote{The terminology of such data can be called ``golden record'', ``gold data'' or even ``gold standard'', ``learning data'' according to the use.}.~\cite{Am2} have been working under the ``no prior knowledge'' hypothesis and managed to calculate the degree of accuracy and precision, assuming that the majority is always right. They defined this degree using the distance of~\cite{Jousselme} between the response and all the other workers' average answers. 

Moreover,~\cite{Dawid} and~\cite{ipr} have been using the Expectation-Maximization (EM) algorithm to estimate the correct response for each task in a first phase which uses labels assigned by the workers. Then, they evaluated the quality of the workers by comparing the responses to the correct inferred answer.

~\cite{smyth} and \cite{raykar} also used this approach for binary classifications and categorical labeling. \cite{raykar2} have generalized this technique on ordinary rankings (associating scores from 1 to 5 depending on the quality of an object or a service). These methods converge to calculate the ``sensitivity'' (the true positives) and the ``specificity'' (the true negatives) for each label. The worker is then labeled as a spammer when his score is closed to 0; A perfect expert would be assigned a score of~1.
The algorithms described previously provide efficient methods to determine the quality of the workers' answers when the truth is unknown whereas in our case the theoretical correct grades attributed to the $R_{i}$ reference signals are known. 
We therefore seek to identify the experts based on correct baseline data and to define a level of expertise proportional to the similarity between worker's answers and known answers in advance. Thus, our work is based on ``golden data'' that are used to estimate the quality of workers in a direct way, as proposed by~\cite{le}. 

When working with ``golden data'', we have the advantage of explicitly measuring the accuracy of workers. The data can be used to make decisions about the workers to check if they are reliable. Can we exploit their results? Should we let them finish the task? Do they deserve a bonus? Additionally, we can also ensure that workers understand completely the nuances and subtleties of the tasks they have to perform. This might be defined as a fully transparent process.

In order to evaluate the impact of using ``golden data'', \cite{ipr} examined the performance of a modified algorithm of~\cite{Dawid} that integrates such type of data. They tried to measure the classification error obtained when varying the percentage of ``golden data'' (0\%, 25\%, 50\% and 75\%). On the one hand the classification error is linked to what extent the algorithm determines the correct class of the examples. On the other hand, the quality estimation error highlights the quality of workers. They found that there is no significant difference between this kind of data and the unsupervised model. Furthermore, they concluded that it is necessary to use ``golden data'' in specific cases as on very imbalanced data sets to evaluate all classes. According to~\cite{ipr}, the most important reasons are the confidence gain of non-technical people (by proposing a quality control approach) and the calibration of results when the emotion-level of the users have an influence on their responses.

\section{Expertise measure proposed}
\label{sec:4}

In this study, to measure the expertise degree we propose a method  based on a comparison of graphs. To evaluate the relevance of this method, we make a comparison with the Fagin distance \cite{Fagin04}, a generalization of the Kendall metric \cite{kendall}, used to count discordant pairs between two ranking lists.

For both of these methods, we consider two types of information:
\begin{enumerate}
	\item The expected theoretical notes which are the correct quality scores, from 1 to 5, associated to the $R_{i}$ reference signals presented in Table~\ref{tab_golddata}.
	
\begin{table}
	\centering
	\begin{tabular}{|c|c|c|c|c|c|}
		\hline
		\textbf{$R_{i}$ reference signals}&$R_{1}$&$R_{2}$&$R_{3}$&$R_{4}$&$R_{5}$\\ \hline
		\textbf{$Sc_{GD}$ ``golden data'' known scores}&1&2&3&4&5\\ \hline
	\end{tabular}
	\caption{``golden data'': $R_{i}$ reference signals associated to their known scores.}
	\label{tab_golddata}
\end{table}
	
	\item The scores attributed by a worker $w$ to the $R_{i}$ reference signals. An example is presented in Table~\ref{tab_workerscores}.

\begin{table}[!h]
	\centering
	\begin{tabular}{|c|c|c|c|c|c|}
		\hline
		\textbf{$R_{i}$ reference signals}&$R_{1}$&$R_{2}$&$R_{3}$&$R_{4}$&$R_{5}$\\ \hline
		\textbf{$Sc_w$ worker related scores}&2&1&2&4&5\\ \hline
	\end{tabular}
	\caption{Example of worker scores on the $R_{i}$ reference signals.}
	\label{tab_workerscores}
\end{table}

\end{enumerate}

\subsection{Belief graph distance-based expertise measure} 
\label{subsec:4_1}

In the proposed method, the answers of the workers on the platform are represented by using oriented and weighted graphs. Then, they are compared to the reference graph constructed on the base of ``golden data''. 

\subsubsection{Graph construction method}
\label{subsubsec:4_1_1}

Graphs are designed as follows:

\begin{itemize}
	\item First a virtual starting point $D$ is inserted with an associated score of 5 (such as the highest score of the $R_{5}$ reference signal).
		
	\item Then, at each iteration $k$, we look for the $R_{i}$ reference signals with the $k^{th}$ highest scores. These $R_{i}$, form the new nodes added to the graph at the same $k^{th}$ depth. The arc ponderation value is equal to the difference between the score associated to the previous nodes and the score of the new nodes added. 
\end{itemize}

\begin{figure}
	\centering
	\includegraphics[width=1\textwidth]{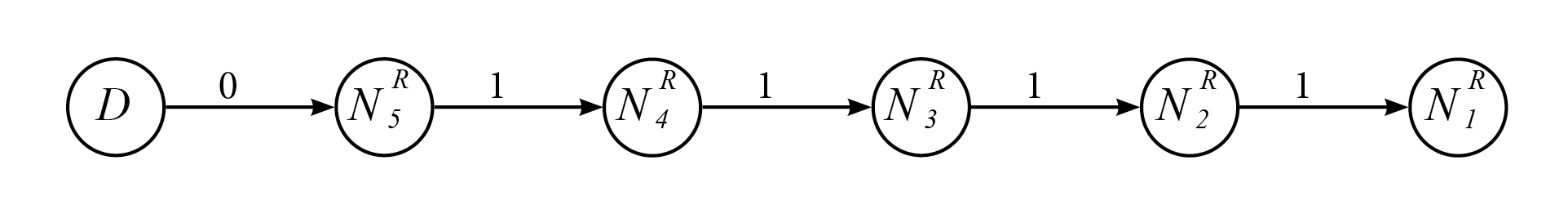}
	\caption{Reference graph constructed with the ``golden data'' known scores.}
	\label{graph_reference}
\end{figure}

By going through this process, the Fig.~\ref{graph_reference} represents the reference graph $G_R$ which corresponds to the $R_{i}$ reference signals associated to their known scores given in Table~\ref{tab_golddata}.

\begin{figure}[!h]
	\centering
	\includegraphics[width=0.95\textwidth]{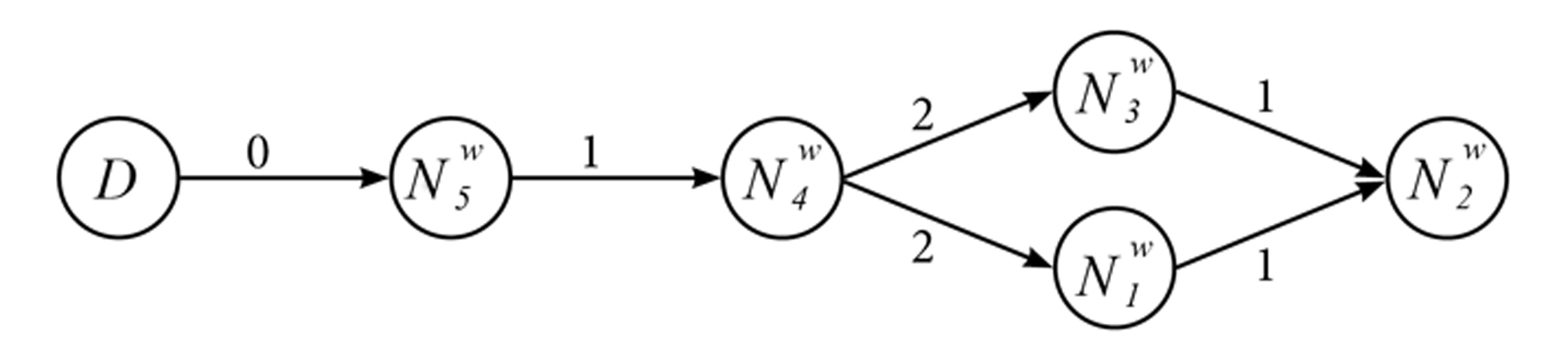}
	\caption{Graph constructed with the worker scores on the $R_{i}$ reference signals.} 
	\label{graph_workerscores}
\end{figure}

In Fig.~\ref{graph_workerscores}, is then presented the graph $G_w$ built on the basis of the worker scores on the $R_{i}$ reference signals given in Table~\ref{tab_workerscores}.

We note $N_i^R$, $N_i^W$ with $i=1,\ldots 5$ respectively the nodes of the reference graph $G_R$ and the nodes of the worker graph $G_w$. We can notice that the nodes $N_i^R$ and $N_i^W$ have the same attribute $R_{i}$ corresponding to the referenced signal. The graph orientation represents the preference order of the $R_{i}$ reference signals. $N_{j}^W \rightarrow N_{k}^W$ means that the score on $R_{j}$ is higher than the score on $R_{k}$. From this order we can define the set of successor nodes $Succ_G(N)$ and the set of predecessor nodes $Pred_G(N)$ $N$ in the graph $G$. 
For example, we have: $Pred_G(N_4^W)=\{N_5^W\}$ and  $Succ_G(N_4^W)=\{N_3^W, N_1^W\}$.

\subsubsection{Expertise degree computation}
\label{subsubsec:4_1_2}

In order to calculate the expertise degrees, the graphs corresponding to the workers' answers are compared with the reference graph and a mass function is thus calculated for each response of the workers.

The objective is to extract a set of heterogeneous factors which enables us to identify the differences between two nodes. This concept is close to the notion of ``signature of the nodes'' introduced by~\cite{Jouili11}, even if the factors considered are not the same.

This model is specific in a way that all the graphs have the same number of nodes with the same labels and the same attributes ({\em i.e.} $R_{i}$). 
According to this method, we need to compare all couples $(N_i^R,N_i^W)$ where $N_i^R$ is the node with the attribute $R_i$ in the reference graph and $N_i^W$ is the node of the same attribute which belongs to the worker graph to be compared.

To do so, we have characterized each node of the graph by using four factors that represent the different possible errors we have identified. These factos are represented and then merged using the mass functions. The discernment frame used is:

\begin{equation}
	\label{notrecadisc}
	\Omega=\{E,NE\}
\end{equation}
where $E$ stands for the Expert's assertion and $NE$ the Non Expert's. We want to measure the belief in the fact that a worker is an expert according to the scores he/she has assigned on the $R_{i}$ reference signals that should be in the correct order.

We describe below the four factors.
\begin{enumerate}
\item \textbf{Accuracy degree of associated scores:} This factor is characterized by the difference of position of a node between the reference graph $G_R$ and the worker's answer graph $G_w$. Dissimilarity is calculated using the Euclidean distance in:

\begin{equation}
	\label{d1}
	d_{1}(N_i^R,N_i^W)=|d_{G_R}(D,N_i^R)-d_{G_W}(D,N_i^W)|
\end{equation}
where $d_G(D,N_i)$ is the node $N_i$'s depth with respect to the node $D$.

For example, consider the nodes corresponding to the reference signal $R_1$ on Fig.~\ref{graph_reference} and Fig.~\ref{graph_workerscores}, we have $d_{G_R}(D,N_1^R)=5$ and $d_{G_W}(D,N_1^W)=3$, so $d_{1}(N_1^R,N_1^W)=2$.

The mass function corresponding to this factor is given by:

\begin{equation}
	\label{m1}
	\begin{cases}
	m_{1}(N_i^R,N_i^W)(E)=\displaystyle 1-\frac{d_{1}(N_i^R,N_i^W)}{d_{max}}\\
	m_{1}(N_i^R,N_i^W)(NE)=\displaystyle \frac{d_{1}(N_i^R,N_i^W)}{d_{max}}
	\end{cases}
\end{equation}

where $d_{max}$ is the maximum distance between two nodes. Given the fact that the graphs considered represent only 5 scores, $d_{max}=4$. 

Thus we obtain on the previous example:
\begin{equation}
	\begin{cases}
	m_{1}(N_1^R,N_1^W)(E)=1-\frac{2}{4}\\
	m_{1}(N_1^R,N_1^W)(NE)=\frac{2}{4}
	\end{cases}
\end{equation}

\item \textbf{Degree of confusion between $R_{i}$ reference signals:} This factor measures the proportion of nodes having the same
distance to the starting point $D$ as the concerned node. Jaccard's dissimilarity will thus be used for the comparison of the set's contents:

\begin{equation}
	\label{d2}
	d_{2}(N_i^R,N_i^W)=\displaystyle \frac{|I_{N_i^R}\bigcap I_{N_i^W}|}{|I_{N_i^R}\bigcup I_{N_i^W}|} 
\end{equation}
where $I_{N_i}=\{R_j; N_j \in V,d_G(D,N_j)=d_G(D,N_i)\}$, with $V$ being the set of nodes of the graph.

If we still consider the nodes corresponding to the reference signal $R_1$ on Fig.~\ref{graph_reference} and Fig.~\ref{graph_workerscores}, we have $I_{N_1^R}=\{R_1\}$ and $I_{N_1^W}=\{R_1, R_3\}$. So we obtain:\\ $d_{2}(N_1^R,N_1^W)=\frac{1}{2}$.

The associated mass function is given by:
\begin{equation}
	\label{m2}
	\begin{cases}
	m_{2}(N_i^R,N_i^W)(E)=d_{2}(N_i^R,N_i^W)\\
	m_{2}(N_i^R,N_i^W)(NE)=1-d_{2}(N_i^R,N_i^W)
	\end{cases}
\end{equation}
The minimum of this mass value is $0.2$.

\item[3-4] {\textbf{Degree of previous bad order} (on the set of predecessors) and \textbf{degree of following bad order} (on the set of successors). Contrary to what is expected (unexpectedly in this precise situation), the worker might consider that a sequence is better than another. Thus, these factors measure these inversion errors according to the previous or the following ones. In order to precise these degrees, we introduce the definition of the following sets, respectively for the set of predecessors (Correct $P_{N_i}^C$ and Non Correct $P_{N_i}^{NC}$) and the set of successors (Correct $S_{N_i}^C$ and Non Correct $S_{N_i}^{NC}$):

\begin{eqnarray}
	\left\{
	\begin{array}{l}
	P^C_{N_i^W}=\{R_j; R_j \in P_{N_i^W}, R_j \in P_{N_i^R}\}\\
	P^{NC}_{N_i^W}=\{R_j; R_j \in P_{N_i^W}, R_j \in S_{N_i^R}\}
	\end{array}
	\right.
\end{eqnarray}
	and
\begin{eqnarray}
	\left\{
	\begin{array}{l}
	S^C_{N_i^W}=\{R_j; R_j \in S_{N_i^W}, R_j \in S_{N_i^R}\}\\
	S^{NC}_{N_i^W}=\{R_j; R_j \in S_{N_i^W}, R_j \in P_{N_i^R}\}
	\end{array}
	\right.
\end{eqnarray}
where $P_{N_i}=\{R_j; N_j \in Pred_{G}(N_i)\}$, $S_{N_i}=\{R_j; N_j \in Succ_{G}(N_i)\}$, with $Succ_G(N)$ and $Pred_G(N)$ are respectively the set of successors and the set of predecessors of the node $N$ in the graph $G$. 

If we consider the nodes corresponding to the reference signals $R_2$, $R_4$ on Fig.~\ref{graph_reference} and Fig.~\ref{graph_workerscores}, we have:
\begin{eqnarray*}
  \begin{array}{ll}
	\left\{
	\begin{array}{l}
	P^C_{N_2^W}=\{R_3\}\\
	P^{NC}_{N_2^W}=\{R_1\}\\
	\end{array}
	\right.
	&
	\left\{
	\begin{array}{l}
	S^C_{N_2^W}=\emptyset\\
	S^{NC}_{N_2^W}=\emptyset
	\end{array}
	\right.
  \end{array}
\end{eqnarray*}
	and
\begin{eqnarray*}
  \begin{array}{cc}
	\left\{
	\begin{array}{l}
	P^C_{N_4^W}=\{R_5\}\\
	P^{NC}_{N_4^W}=\emptyset
	\end{array}
	\right.
	&
	\left\{
	\begin{array}{l}
	S^C_{N_4^W}=\{R_3\}\\
	S^{NC}_{N_4^W}=\emptyset
	\end{array}
	\right.
  \end{array}
\end{eqnarray*}

From these definitions, the distances $d_3$ and $d_4$ are given by the following equations:
\begin{eqnarray}
	\label{d3}
	\left\{
	\begin{array}{l}
	d_{3,1}(N_i^R,N_i^W)=\displaystyle \frac{|P^C_{N_i^W}\bigcap P_{N_i^R} |}{|P_{N_i^R} \bigcup P_{N_i^W} |}=m_{3}(N_i^R,N_i^W)(E)\\
	d_{3,2}(N_i^R,N_i^W)=\displaystyle \frac{|P^{NC}_{N_i^W} |}{|P_{N_i^W}|}=m_{3}(N_i^R,N_i^W)(NE)
	\end{array}
	\right.
\end{eqnarray}

\begin{eqnarray}
	\label{d4}
	\left\{
	\begin{array}{l}
	d_{4,1}(N_i^R,N_i^W)=\displaystyle \frac{|S^C_{N_i^W}\bigcap S_{N_i^R} |}{|S_{N_i^R} \bigcup S_{N_i^W} |}=m_{4}(N_i^R,N_i^W)(E)\\
	d_{4,2}(N_i^R,N_i^W)=\displaystyle \frac{|S^{NC}_{N_i^W} |}{|S_{N_i^W}|}=m_{4}(N_i^R,N_i^W)(NE)
	\end{array}
	\right.
\end{eqnarray}
	
The rest of the mass will be used to weigh ignorance. The mass associated with ignorance can also be derived from extreme nodes that are without predecessors (all nodes except node (5)) or successors (all nodes except node (1)).

Thus we obtain on the example:
\begin{eqnarray}
	\begin{array}{ll}
	\left\{
	\begin{array}{l}
	d_{3,1}(N_2^R,N_2^W)=1/2\\
	d_{3,2}(N_2^R,N_2^W)= 1/2
	\end{array}
	\right.
	&
	\left\{
	\begin{array}{l}
	d_{4,1}(N_2^R,N_2^W)=0\\
	d_{4,2}(N_2^R,N_2^W)=0
	\end{array}
	\right.
	\end{array}
\end{eqnarray}

Here, we consider that $0/0$ is 0. 
\begin{eqnarray}
	\begin{array}{ll}
	\left\{
	\begin{array}{l}
	d_{3,1}(N_4^R,N_4^W)=1\\
	d_{3,2}(N_4^R,N_4^W)=0
	\end{array}
	\right.
	\left\{
	\begin{array}{l}
	d_{4,1}(N_4^R,N_4^W)= 1/2\\
	d_{4,2}(N_4^R,N_4^W)=0
	\end{array}
	\right.
	&
	\end{array}
\end{eqnarray}

Equations (\ref{d1}), (\ref{m1}), (\ref{d2}), (\ref{m2}), (\ref{d3}) and (\ref{d4}) provide a way to calculate the mass functions by using a set of factors for each pair of nodes $(N_i^R,N_i^W)$ according to the reference graph $G_R$ and the graph $G_w$ which corresponds to the worker answers with attribute $i$. The next step defines a mass function on the entire graph by averaging the mass functions on all the nodes, calculated for each factor:
	
\begin{eqnarray}
	\label{mbba_crit}
	\left\{
	\begin{array}{l}
	m_{k}(G_{R},G_{W})(E)=\displaystyle \frac{\displaystyle \sum_{i=1}^{O(G)}m_{k}(N_i^R,N_i^W)(E)}{O(G)} \\
	m_{k}(G_{R},G_{W})(NE)=\displaystyle \frac{\displaystyle \sum_{i=1}^{O(G)}m_{k}(N_i^R,N_i^W)(NE)}{O(G)}
	\end{array}
	\right.
\end{eqnarray}	
where $O(G)$ is the graph's order ({\em i.e.} the number of vertices, here 6).


In order to obtain a mass function for the considered response, we combine the mass functions of the four factors. Finally, the degree of expertise is given by calculating the distance from~\cite{Jousselme} between the mass function and the categorical mass function on the expert element such as \linebreak\cite{Essaid2014}.}
\end{enumerate}

\subsection{Expertise degree based on the Fagin distance}
\label{subsec:4_2}

To compare two ranking lists, a well-known way is to use the metric defined by~\cite{kendall} that applies a penalty when different orders are encountered in the two rankings. A generalization of this distance has been proposed by \linebreak \cite{Fagin04} in case of partial rankings. A first step is based on the Kendall metric with the definition of the $p$ penalty in the $[0, 1]$ interval, and a second step based on the Hausdorff distance. We detail both steps bellow. 

Both lists considered here are given by the corresponding scores of the reference signals $R_i$: 
\begin{itemize}
	\item $Sc_{GD}$ containing the ``golden data'' known scores according to the reference signals $R_i$  ({\em cf.} Table~\ref{tab_golddata}, {\em e.g.} $Sc_{GD}(R_3)=3$)
	\item $Sc_{w}$ containing the scores proposed by the worker $w$ on one HIT 
	according to the reference signals $R_i$ ({\em cf.} Table~\ref{tab_workerscores}, {\em e.g.} $Sc_{w}(R_3)=2$)
\end{itemize}

The Kendall distance $K^{(p)}(Sc_{GD}, Sc_w)$ between these two lists is defined by: 

\begin{equation}
	\label{kendall}
	K^{(p)}(Sc_{GD}, Sc_w) = \sum_{\{i,j\}\in P} \bar K^{(p)}_{i,j}(Sc_{GD}, Sc_w) 
\end{equation}
where $P$ is the set of unordered pairs of distinct elements in $Sc_{GD}$ and $Sc_w$, and $i < j$. Therefore we have: $Sc_w(R_i)<Sc_w(R_j)$.

Two cases are taken into account to determine the value of $p$:
\begin{itemize}
	\item case 1: in both lists, $Sc_{GD}$ and $Sc_w$, $i$ and $j$ are in different buckets\footnote{A bucket is a set of musical sequences with the same score. In the $Sc_{GD}$ list all the five musical sequences are in different buckets.}:
	\begin{itemize}
		\item if the order is the same for $i$ and $j$ in the two lists, then $\bar K^{(p)}_{i,j}(Sc_{GD}, Sc_w)=0$. $Sc_w(R_i) < Sc_w(R_j)$ as $Sc_{GD}(R_i) < Sc_{GD}(R_j)$ induces that there is no penalty for ${\{i,j\}}$. That is the case on the previous example given by Table~\ref{tab_workerscores} for ${\{4,5\}}$, $\bar K^{(p)}_{4,5}(Sc_{GD}, Sc_w)=0$.
		\item if the order is different for $i$ and $j$ in the two lists, then $\bar K^{(p)}_{i,j}(Sc_{GD}, Sc_w)=1$. $Sc_w(R_i) > Sc_w(R_j)$ unlike $Sc_{GD}(R_i) < Sc_{GD}(R_j)$ induces that the penalty for ${\{i,j\}}$ is equal to $1$. That is the case on the previous example given by Table~\ref{tab_workerscores} for ${\{1,2\}}$, $\bar K^{(p)}_{1,2}(Sc_{GD}, Sc_w)=1$.
	\end{itemize}
	\item case 2: in the $Sc_w$ list, $i$ and $j$ are in the same bucket, whereas they are in different buckets in the $Sc_{GD}$ list, then $\bar K^{(p)}_{i,j}(Sc_{GD}, Sc_w)=p$. The value $p$ must be between 0.5 and 1 in order to obtain a distance. In the rest of the paper we choose $p=0.5$, but the results do not change a lot for higher values. That is the case on the previous example given by Table~\ref{tab_workerscores} for ${\{1,3\}}$, $\bar K^{(p)}_{1,3}(Sc_{GD}, Sc_w)=0.5$.
\end{itemize}

We can notice that a third case (where $i$ and $j$ would be in the same buckets in both lists) can not occur here because ``golden data'' are all different.

In a second step, we consider the answers of a worker for more than one HIT. 
We consider the two sets of list scores of corresponding HIT given by $H_w$, $H_R$.~\cite{Fagin04} propose to consider the Hausdorff metrics between two objects $H_w$, $H_R$ given by:
	\begin{equation*} 
		d_{Haus}(H_w,H_R)=\max \left( \max_{Sc_w \in H_w} \min_{Sc_{GD} \in H_R} \, d(Sc_w,Sc_{GD}), \max_{Sc_{GD} \in H_R} \min_{Sc_w \in H_w} \, d(Sc_w,Sc_{GD}) \right)
	\end{equation*}
The distance $d(Sc_w,Sc_{GD})$ is given by the equation~\eqref{kendall}. This distance can be used as an expertise measure with values into $[0, 1]$.

\subsection{Comparison on a simple example}
\label{subsec:4_3}

Even if it has been adapted to integrate ties, the Fagin-based expertise measure principle gives importance to the order concordance between the two considered lists. In the proposed belief-based measure, the focus is more on the difference between the two list scores. Taking into account these characteristics, the expertise degree obtained for specific types of responses may be unsatisfactory.
For example, on table~\ref{ExampleAns} we show an answer for a worker that does not make the difference between the musical sequences or does not want to answer with honesty. In that case the belief based expertise measure gives a low value of 0.196, leading to the non selection of the worker. However, the Fagin-based expertise measure gives a value of 0.65 suggesting the worker is an expert. 


\begin{table}[!h]
	\centering
	\begin{tabular}{|c|c|c|c|c|c||c|c|c|c|c||c|c|c|c|c||c|c|c|c|c|}
		\hline
		& \multicolumn{5}{c||}{HIT 1}& \multicolumn{5}{c||}{HIT 2}& \multicolumn{5}{c||}{HIT 3}& \multicolumn{5}{c|}{HIT 4} \\
		\hline
		\textbf{$R_{i}$ reference signals}&1&2&3&4&5&1&2&3&4&5&1&2&3&4&5&1&2&3&4&5\\ \hline
		\textbf{$Sc_w$ worker related non-pertinent scores}&5&5&5&5&5&5&5&5&5&5&5&5&5&5&5&5&5&5&5&5\\ \hline
	\end{tabular}
	\caption{Example of some answers of a worker on four HITs.}
	\label{ExampleAns}
\end{table}

Of course, the definition of expert from such measures must be made according to a threshold. This threshold can be defined according to the experimental values as presented in the next section.

\section{Evaluation of methods in real situation}
\label{sec:5}

Historically, Orange Labs has been performing subjective testing of audio coders in the laboratory. These tests consist in recruiting listeners deemed to be naive which means they are not directly involved in work related to quality evaluation or audio coding. Short speech or music sequences processed according to different coders are presented to these workers to allow them to evaluate the audio quality on suitable scales.

The tests take place in acoustically treated rooms and more generally in a perfectly controlled environment. These laboratory methods are effective but still costly and might lead to results with a low representativity (relative to the use of in situ services) or limited stimuli (limited number, for example).

In order to add the crowdsourcing approach to the test methods, two campaigns were implemented on a crowdsourcing platform and the results were compared to those obtained in the laboratory. Each campaign was a replica of the same test initially performed in the laboratory for the G729EV coder's standardization.

In this laboratory test, 7 test conditions (\textit{ i.e.} coding solutions) were tested, to which were added the 5 $R_{i}$ reference conditions (MNRUs). In total, 12 conditions have been tested through 12 musical sequences presented in random order. These sequences constituted a HIT (\textit{Human Intelligence Task}). 32 people participated to the test which falls into 4 groups. Each group listened to 4 HITs and evaluated each of them. After each audio sequence, listeners were asked to rate the quality on a scale of 1 (= Bad) to 5 (= Excellent).

As in the laboratory experimentation, workers in the crowdsourcing campaigns, were divided into 4 distinct panels (each worker can only belongs to one panel). According to the laboratory experimental test design, each panel was given 4 HITs of 12 audio sequences to be evaluated on the same quality scale. Each HIT was related to 1 micro-job on the crowdsourcing platform. As a result, each worker could provide a contribution from 1 to 4 micro-jobs (the 4 HITs of his panel are different from those of the other panels). Each participation was taken into account if the worker had completed at least one HIT. Unlike what is practiced in the laboratory environment, the worker could stop listening before the end of the HIT. Instructions were presented in English writing to the workers before the test. A training session with 8 audio sequences was also performed before the test, as in the laboratory.

Two campaigns were carried out on two different geographical areas. All the English speaking workers were allowed to participate in the first campaign, regardless of their country. Workers who took part in this campaign were mostly located in Asia. The second campaign was limited to the USA. Both campaigns were carried out using the same conditions (workers belonging to the same $Panel_i$ and listening to the same sequences for the two campaigns).

\subsection{Analysis of belief graph distance-based expertise measure}
\label{subsec:5_1}
\begin{figure}[!h]
	\includegraphics[width=0.9\textwidth]{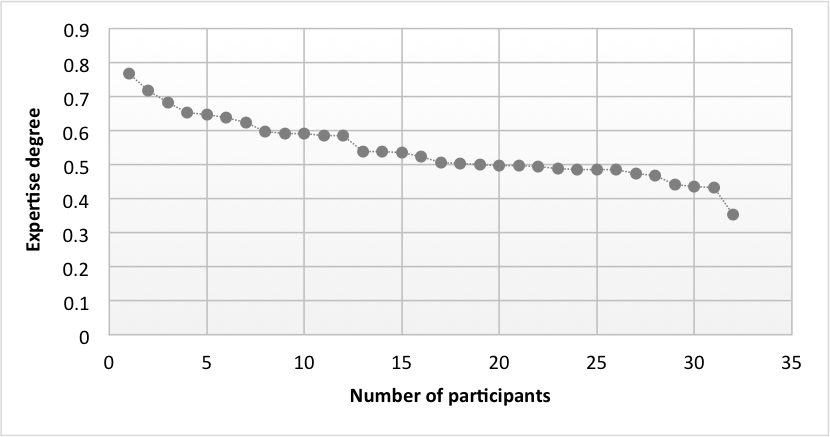}
	\caption{Expertise degree of laboratory workers with the belief-based measure.}
	\label{ExpDegLabBG}
\end{figure}

\begin{figure}[!h]
	\includegraphics[width=0.9\textwidth]{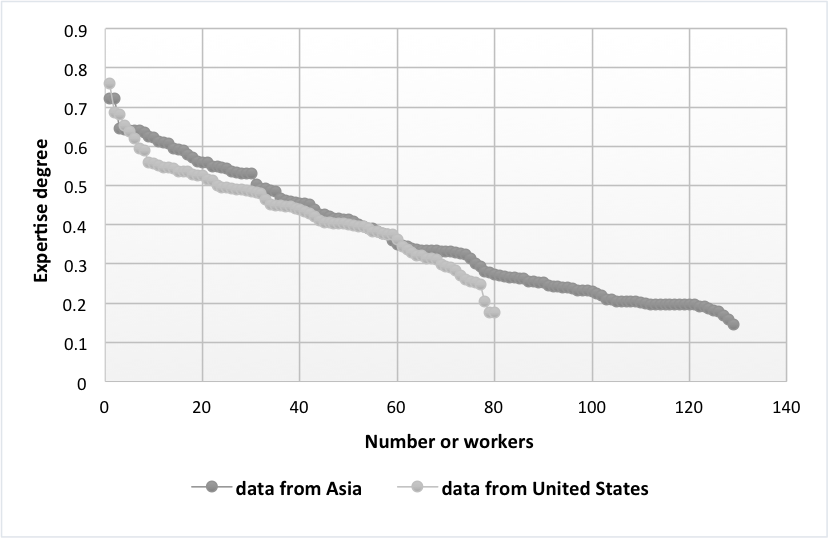}
	\caption{Expertise degree of crowdsourcing workers with the belief-based measure.}
	\label{ExpDegCwdBG}
\end{figure}

The expertise degrees based on belief functions were calculated using the laboratory data (Fig.~\ref{ExpDegLabBG}) on the one hand, and using the scores from the crowdsourcing platforms (Fig.~\ref{ExpDegCwdBG}) on the other hand.

31 people out of 32 obtained an expertise degree greater than 0.4 (this threshold chosen in comparison with data from the platform). These results highlight the reliability of the answers collected in the laboratory. The interval $[0.4, 0.5]$ contains the largest number of workers.

Firstly, the expertise degree distributions calculated on the 4 panels of crowdsourcing platforms are illustrated on Fig.~\ref{ExpDegCwdBG}. When looking at the distributions in question, one can notice a small gap on the interval $[0.4, 0.5]$. This allows us to determine the most discriminating threshold of expertise ({\em i.e.} the threshold which allows to split the workers into 2 distinct groups).

We observe that the computed expertise degrees vary in a wider range than in the laboratory experimentation (from 0.1 to 0.7). The explanation to this phenomenon can be articulated in two directions:
\begin{itemize}
	\item The first factor could be the lack of seriousness of a larger number of workers on the crowdsourcing platform;
	\item The listening conditions ({\em e.g.} sound environment, headphones or loudspeaker (s), PC used) vary from one worker to another, from one HIT to another, unlike the laboratory situation, and might have influenced the quality of workers' answers.
\end{itemize}
Our objective in this work is not to impose strict experimentation conditions but to put workers in a familiar context. Moreover, by comparing the two distributions, we notice a small difference between the two campaigns. For example, the interval of expertise $[0.1, 0.2]$ is almost absent for the US campaign data (2 people on all panels, see gray curve), whereas the same interval contains 19 workers in the Asian campaign (see black curve). Furthermore, for Asia, the interval $[0.2, 0.3]$ contains most of the workers, whereas for the USA, the interval $[0.4, 0.5]$ is preponderant. The differences observed between the two campaigns can be explained by the cultural differences between the two regions. In particular, the American workers are culturally closer to selected musical sequences (occidental music).

In a first analysis, we select $0.4$ as the threshold because it is closer to the gaps in the distributions (Fig.~\ref{ExpDegCwdBG}). We select the workers with an expertise degree over this threshold: thus we keep 51 workers out of 129 for the first campaign and 50 workers out of 80 for the second one. Hence we suppress more workers for the first campaign. 

The average of their answers will be taken into account for the evaluation of audio quality. We compare the data from the two campaigns on the crowdsourcing platforms with those obtained in the laboratory according to this threshold.

\subsection{Analysis of Fagin distance-based expertise measure}
\label{subsec:5_2}

\begin{figure}
	\includegraphics[width=0.9\textwidth]{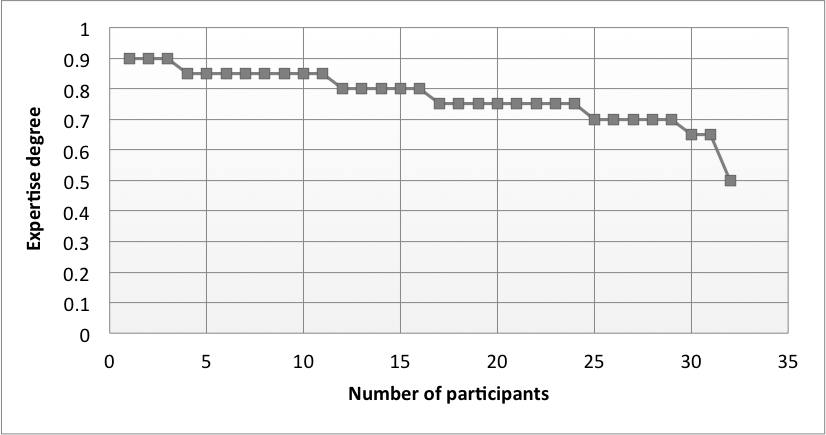}
	\caption{Expertise degree of laboratory workers with the Fagin distance.}
	\label{ExpDegLabKen}
\end{figure}

\begin{figure}
	\includegraphics[width=0.9\textwidth]{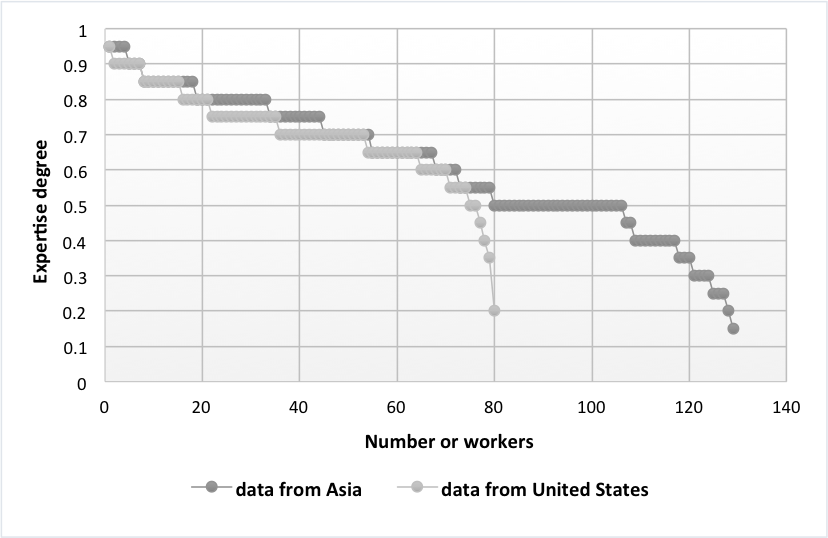}
	\caption{Expertise degree of crowdsourcing workers with the Fagin distance.}
	\label{ExpDegCwdKen}
\end{figure}

As in the previous method we have already mentioned, the expertise degrees based on Fagin distance were calculated using the laboratory data (Fig.~\ref{ExpDegLabKen}) on the one hand, and using the scores from the crowdsourcing platforms (Fig.~\ref{ExpDegCwdKen}) on the other hand. 

We can first notice a higher range of value compared to the belief-based measure. Considering the two curves, we select $0.6$ as the threshold which corresponds to a gap in the distributions, especially in the Asian one.

Here again for the laboratory experimentation, we obtain 31 people out of 32 over this expertise degree (Fig.~\ref{ExpDegLabKen}). As a matter of fact, the worker with the lowest expertise degree is the same as the non-selected worker with the belief-based measure. 

According to Fig.~\ref{ExpDegCwdKen} which presents the results for the crowdsourcing campaigns, we can notice that the Fagin-based measure cannot discriminate some workers. Indeed, the Kendall distance definition, in the case of a comparison of 5 values, gives expertise degrees with a precision of only 0.05. 

Considering the $0.6$ threshold, with this measure we keep 67 workers out of 129 workers for the first campaign and 64 workers out of 80 workers for the second one. Hence we discard more workers for the first campaign.

\subsection{Fusion of belief-based and Fagin distance-based expertise measures}
\label{subsec:5_3}
As we show in section~\ref{subsec:4_3}, when we compare the belief-based measure and the Fagin distance, one can consider a worker such as an expert and the other not. Consequently, a fusion of these two methods is interesting, especially in the crowdsourcing approach in order to select workers both qualified as experts with the two measures. 

Therefore, we only take into account the workers whose expertise degree is greater than 0.4 with the belief approach and greater than 0.6 with the Fagin distance. Hence, the objective of the fusion of belief-based measure and Fagin distance is to reduce errors and to ensure the results when we detect experts. Thus, we obtain 47 workers considered as experts in the first campaign and 49 workers in the second one. The number of selected workers for laboratory data is still 31.
Even if the number of workers on the first campaign (129) is greater than on the second one (80), the number of selected experts on the second campaign is greater. 

\begin{figure}
	\includegraphics[width=1\textwidth]{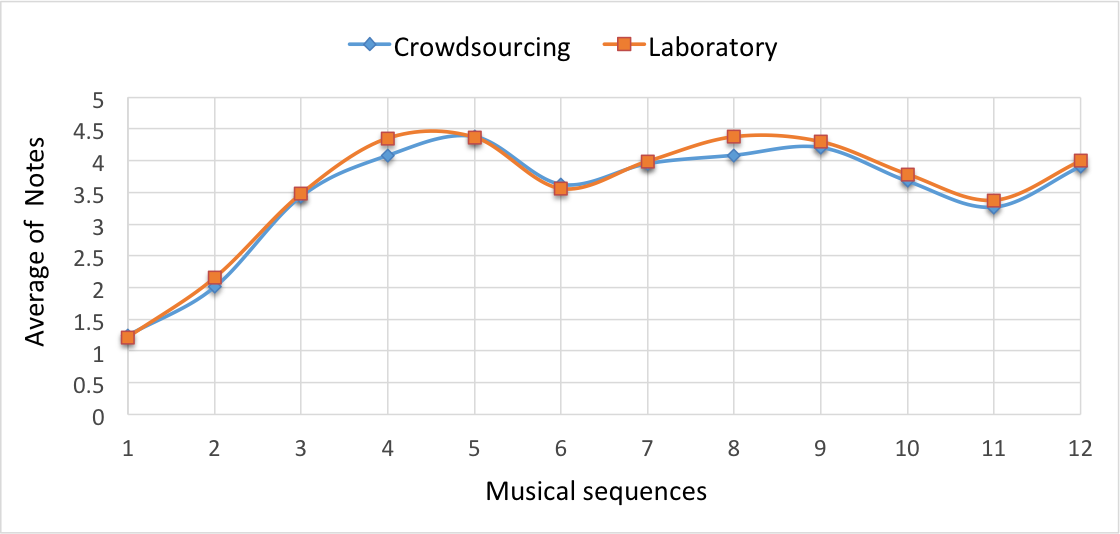}
	\caption{Comparison of the laboratory / crowdsourcing average of the notes given on the 12 musical sequences for the first campaign (Asia).}
	\label{compAsia}
\end{figure}

\begin{figure}
	\includegraphics[width=1\textwidth]{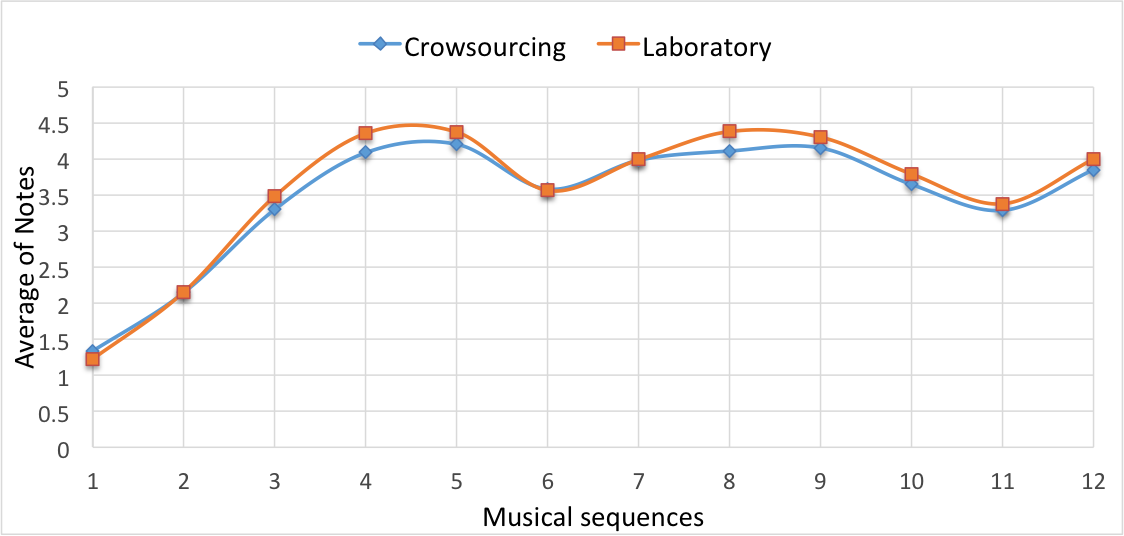} 
	\caption{Comparison of the laboratory / crowdsourcing average of the notes given on the 12 musical sequences for the second campaign (USA).}
	\label{compUSA}
\end{figure}

In order to compare the results given by the selected workers in laboratory and the selected workers on the crowdsourcing platform, we calculate the average of the scores for each of the 12 musical sequences. The first 5 musical sequences are the $R_{i}$ reference signals. Hence, the expected scores are 1, 2, 3, 4 and 5. We notice that the laboratory and the crowdsourcing curves for both campaigns (see figures~\ref{compAsia} and~\ref{compUSA}) are very close. In an optimal situation, the curves linked to the first five sequences would be a straight line as they correspond to the $R_{i}$ reference signals. However, these results are explained by the usual behaviors that we observe here on laboratory and platform data where workers are reluctant to give the maximum score of 5. 

Finally, the proximity between the two curves emphasizes the benefit of carrying out this type of evaluation on crowdsourcing platforms once the workers with the highest degrees of expertise have been selected. 
In view of these positive achievements, we can conclude that the experts have been selected efficiently.

\section{Conclusion and discussion}
\label{sec:6}

In this work we propose an innovative approach to compute the expertise of workers in a subjective evaluation of audio quality through listening tests. This approach is based on the modeling of the workers' scores through an oriented graph. Taking into account the data whose expected order of preference is known, we have developed a measure of comparison of two graphs. Thus, the approach is based on four factors from which four mass functions have been defined in order to account for the possible imperfections of workers' answers. From these mass functions, a level of expertise is computed for each worker so that only workers with a sufficient level of expertise can be considered.

Moreover, to evaluate the relevance of this method, we make a comparison with the Fagin distance. Some differences on the characterization of experts lead us to propose a fusion of the two measures.

The comparison of data providing from crowdsourcing and laboratory campaigns prove the benefit of conducting such evaluations from crowdsourcing platforms. However, it is necessary to accurately evaluate the workers' degree of expertise in order to exclude irrelevant answers coming from crowdsourcing platforms. Thanks to the approach developed in this work for the evaluation of the expertise degrees, we are now in position to exclude the workers with no relevant answers in the task of audio quality evaluation.

 \bibliographystyle{apalike}
 \bibliography{refdis} 
%


\end{document}